# Model Predictive Approach for Detumbling an Underactuated Satellite

Kota Kondo[1], Yasuhiro Yoshimura[2], Mai Bando[3],

Shuji Nagasaki[4], Toshiya Hanada[5]

*Kyushu University, Fukuoka-shi, Fukuoka, 819-0395, Japan*

This research proposes an innovative approach to detumble satellites' triple-axis angular velocities with only one single-axis magnetic torquer. Since magnetic torque is generated perpendicularly to magnetorquers, no intended control torque along the magnetorquer can be produced, which makes systems underactuated. Our paper introduces a control method using Model Predictive Control (MPC) and compares it with B-dot control algorithm. By applying these control laws to Kyushu University Light Curve Inversion (Q-Li) Demonstration Satellite in numerical simulations, we describe the applicability of these control laws to underactuated systems.

## I. Nomenclature

| | | |
|---|---|---|
| $\boldsymbol{B}$ | = | Earth magnetic field vector |
| $H$ | = | Hamiltonian |
| $h$ | = | a small constant |
| $\boldsymbol{J}$ | = | moment of inertia |
| $\boldsymbol{m}$ | = | magnetic dipole vector |
| $m_{\max}$ | = | maximum input of magnetorquer |
| $m_x$ | = | optimal input on $x$ axis |
| $N$ | = | discretization maximum step number |
| $Q, R_1, R_2$ | = | optimal weight function |
| $T$ | = | final time on horizon |
| $\boldsymbol{T}$ | = | control torque vector |

---

[1] Student, Department of Mechanical and Aerospace Engineering, AIAA Student Member.
[2] Assistant Professor, Department of Aeronautics and Astronautics.
[3] Associate Professor, Department of Aeronautics and Astronautics.
[4] Assistant Professor, Department of Aeronautics and Astronautics.
[5] Professor, Department of Aeronautics and Astronautics.

| | | |
|---|---|---|
| $t$ | = | initial time on horizon |
| $\boldsymbol{U}(t)$ | = | optimal input matrix |
| $V$ | = | Lyapunov function |
| $v$ | = | dummy input |
| $\boldsymbol{x}$ | = | state variable |
| $\Delta t$ | = | sampling time span |
| $\boldsymbol{\omega}$ | = | angular velocity |
| $\rho$ | = | Lagrange multiplier for equality constraint |
| $\lambda$ | = | Lagrange multiplier |
| $(\ )_i$ | = | $i^{\text{th}}$ time step on MPC method |
| $(\ )^*$ | = | conditions on MPC method |

## II. Introduction

Since the beginning of satellite designs, especially for small satellites, there have been inflexible restrictions on their sizes, weights, budgets; therefore, underactuated systems, which lead to research in minimizing sizes and weights savings, have been long researched [1,2]. An underactuated system is defined as a system which equips fewer actuators than its degrees of freedom [3]. Underactuated control methods can serve as additional layers of redundancy, and theorems have been developed by researchers on this topic.

In recent years, universities have been developing small satellites, and the demand of underactuated systems has been significantly increased. Kyushu University Light Curve Inversion Demonstration Satellite (Q-Li) project is a university project, where students are developing a 3U satellite to verify techniques of light curve inversion and to detect space debris [4]. To achieve the aforementioned objectives, the Q-Li satellite needs to deploy a thin membrane structure, which requires its angular velocities to be sufficiently attenuated. The Q-Li satellite, however, has only one single-axis magnetorquer due to restrictions on its size and budget; therefore, an underactuated controller that achieves 3-axis detumbling with only the single-axis magnetic torquer is in need.

This study investigates the applicability of two magnetic detumbling methods: B-dot algorithm and Model Predictive Control (MPC), a non-linear feedback optimal control method. B-dot algorithm has been applied to stabilize angular rates; nonetheless, since it is designed for 3-axis magnetic torquers, this algorithm is found difficult to stabilize with a single-axis magnetic torquer. In addition, the Q-Li's axisymmetric moment inertia makes it difficult to detumble the angular rate around its symmetry axis. B-dot algorithm is theoretically inapplicable in this case. Yet, by using variable optimal weight functions, MPC overcomes this difficulty. To demonstrate the effectiveness of the MPC method, this paper shows numerical simulation results by applying the control laws to the Q-Li satellite.

## III. Detumbling Control Law

### A. Dynamics

The body fixed frame of a rigid body is assumed to be located at the center of mass and to be aligned with the principal axes of inertia. The equation of a rigid body's rotational motion in a body-fixed frame is described as follows [5,6].

$$\begin{bmatrix} \dot{\omega}_x \\ \dot{\omega}_y \\ \dot{\omega}_z \end{bmatrix} = \begin{bmatrix} \frac{1}{J_x}\{(J_y - J_z)\omega_y\omega_z\} \\ \frac{1}{J_y}\{(J_z - J_x)\omega_z\omega_x\} \\ \frac{1}{J_z}\{(J_x - J_y)\omega_x\omega_y\} \end{bmatrix} + \begin{bmatrix} \frac{T_x}{J_x} \\ \frac{T_y}{J_y} \\ \frac{T_z}{J_z} \end{bmatrix} \quad (1)$$

A single-axis magnetic torquer interacts with the Earth's local magnetic field and generates the control torque vector **T** as given below [7].

$$\boldsymbol{T} = \boldsymbol{m} \times \boldsymbol{B} \quad (2)$$

Given the single-axis magnetic torquer is on the *x*-axis, the control torque with the single-axis magnetorquer is written as follows.

$$\boldsymbol{T} = \begin{bmatrix} 0 \\ -B_z m_x \\ B_y m_x \end{bmatrix} \quad (3)$$

As seen in Eq. (3), no control input torque is generated on the *x*-axis, which makes this control system underactuated. Taking these characteristics into account, the Euler's equation is eventually given as follows.

$$\begin{bmatrix} \dot{\omega}_x \\ \dot{\omega}_y \\ \dot{\omega}_z \end{bmatrix} = \begin{bmatrix} \frac{1}{J_x}\{(J_y - J_z)\omega_y\omega_z\} \\ \frac{1}{J_y}\{(J_z - J_x)\omega_z\omega_x - B_z m_x\} \\ \frac{1}{J_z}\{(J_x - J_y)\omega_x\omega_y + B_y m_x\} \end{bmatrix} \quad (4)$$

The Q-Li satellite is nearly axisymmetric around its *x*-axis; hence, the *y* and *z* axes' moments of inertia are nearly equal, which makes the time derivative of *x*-axis angular rate nearly zero. However, since it is not completely zero, we can still stabilize the *x*-axis velocity even though it is challenging to detumble *x*-axis with only input control $m_x$ from *y* and *z* axis.

**B. B-dot Control Algorithm**

To detumble small satellites with magnetorquers, B-dot algorithm has been long applied. The principle of B-dot algorithm is to reduce satellites' kinetic energy, which leads to a reduction in its angular velocities. We first demonstrate the stability of B-dot algorithm with a 3-axis magnetic torquer. Afterwards, we discuss why we cannot achieve 3-axis detumbling with a single-axis magnetic torquer on B-dot control algorithm. The equation of motion of satellites is given as Eq. (5). (This is a different form of Eq. (1).)

$$J\dot{\boldsymbol{\omega}} + \boldsymbol{\omega} \times J\boldsymbol{\omega} = \boldsymbol{T} \tag{5}$$

As the state vector $\boldsymbol{x} \equiv \boldsymbol{\omega}$, we define Lyapunov function as Eq. (6) [8].

$$V(\boldsymbol{x}) = \frac{1}{2}\boldsymbol{\omega}^T J \boldsymbol{\omega} \tag{6}$$

The Lyapunov function is always positive except when $\boldsymbol{\omega} = 0$ (Equilibrium point). The derivative of it is given in Eq. (7) by using Eq. (5).

$$\begin{aligned} \dot{V}(\boldsymbol{x}) &= \boldsymbol{\omega}^T J \dot{\boldsymbol{\omega}} \\ &= \boldsymbol{\omega}^T(-\boldsymbol{\omega} \times J\boldsymbol{\omega} + \boldsymbol{T}_m) \\ &= \boldsymbol{\omega}^T \boldsymbol{T} \end{aligned} \tag{7}$$

The feedback control input of magnetic dipole moment is given in [9,10]. The $m_{\text{max}}$ is the maximal value that a single axis magnetorquer can generate.

$$\boldsymbol{m} = -m_{\text{max}} \frac{\dot{\boldsymbol{B}}}{\|\dot{\boldsymbol{B}}\|} \tag{8}$$

The time derivative of Earth's magnetic field vector $\boldsymbol{B}$ with respect to an inertial frame is

$$^I\dot{\boldsymbol{B}} = \dot{\boldsymbol{B}} + \boldsymbol{\omega} \times \boldsymbol{B} \tag{9}$$

where the left superscript $I$ on $\boldsymbol{B}$ means "with respect to an inertial frame". Since $^I\dot{\boldsymbol{B}}$ is small enough compared to $\dot{\boldsymbol{B}}$ and the derivative of Earth's magnetic field vector is with respect to a body fixed frame [11,12], We now have

$$0 \approx \dot{\boldsymbol{B}} + \boldsymbol{\omega} \times \boldsymbol{B} \tag{10}$$

$$\rightarrow \dot{B} \approx -\omega \times B \tag{11}$$

Taking all Eqs. (2), (6), and (11) into account, we have $\dot{V}(x)$ as described in Eq. (12).

$$\begin{aligned}
\dot{V}(x) &= \omega^T T \\
&= \omega^T (m \times B) \\
&= -\omega^T (B \times m) \\
&= -(\omega \times B)^T m \\
&= -(\omega \times B)^T \left(-m_{\max} \frac{\dot{B}}{\|\dot{B}\|}\right) \\
&= -(\omega \times B)^T \left(\frac{m_{\max}}{\|\dot{B}\|} \omega \times B\right) \\
&= -\frac{m_{\max}}{\|\dot{B}\|} (\omega \times B)^T (\omega \times B) \\
&= -\frac{m_{\max}}{\|\dot{B}\|} \|\omega \times B\|^2
\end{aligned} \tag{12}$$

$\dot{V}(x) = 0$ when $\omega = 0$. (Technically, $\dot{V}(x) = 0$ when $\omega$ is parallel to $B$ too; however, it is caused an assumption— $^I\dot{B}$ is small enough compared to $\dot{B}$, the derivative of Earth's magnetic field vector with respect to a body fixed frame—.) By LaSalle's invariance principle, $\omega = 0$ is asymptotically stable. This is a basic approach to show why B-dot algorithm can detumble 3-axis angular velocities [13]. Now, in cases of a control system that uses only one single-axis magnetic torquer—an underactuated satellite, we show B-dot algorithm cannot be applied. In that case, Eq. (12) is described as follows.

$$\begin{aligned}
\dot{V}(x) &= -(\omega \times B)^T \cdot m \\
&= -(\omega \times B)^T \cdot \begin{bmatrix} m_x \\ 0 \\ 0 \end{bmatrix} \\
&= -\frac{m_{\max}}{\|\dot{B}\|} (\omega \times B)^T \cdot \begin{bmatrix} \omega_y B_z - \omega_z B_y \\ 0 \\ 0 \end{bmatrix} \\
&= -\frac{m_{\max}}{\|\dot{B}\|} (\omega_y B_z - \omega_z B_y)^2
\end{aligned} \tag{13}$$

Equation (13) also satisfies $\dot{V}(x) \leq 0$ and it does not depend on $\omega_x$. In other words, it is necessary to satisfy $\dot{V}(x) = 0$, $\omega_y = 0$ $and$ $\omega_z = 0$ while $\omega_x$ can be arbitrary. Therefore, Lasalle's largest positively invariant set $N$ is given as the following.

$$N = \left\{ \boldsymbol{\omega} = \begin{bmatrix} arbitrary \\ 0 \\ 0 \end{bmatrix} \middle| \dot{V} = 0 \right\} \tag{14}$$

Equation (14) indicates, in an underactuated system, the state vector $x(t)$ ($\equiv \boldsymbol{\omega}(t)$) will eventually approach the largest positively invariant set, where $\omega_x$ does not necessarily have to be zero. Thus, the underactuated B-dot algorithm cannot be applied to the 3-axis detumbling.

### C. Model Predictive Control

Since the equation of motion of satellites shown in Eq. (2) is non-linear, control methods for linear systems cannot generally be applied to the small satellites' detumbling. Hence, MPC method, a feedback optimal control method, which can be applied to non-linear systems, is suitable in this case. We define the cost function $J$, considering variable optimal weight functions, $Q, R_1, R_2$, 3-axis angular velocities, and dummy input $v$. [14]

$$J = \frac{1}{2}\left(\omega_x^2(t+T) + \omega_y^2(t+T) + \omega_z^2(t+T)\right) \\ + \int_t^{t+T} \left\{ \frac{1}{2}(Q_1\omega_x^2 + Q_2\omega_y^2 + Q_3\omega_z^2) + R_1 m_x^2\right\} - R_2 v \right\} d\tau \tag{15}$$

The dummy input $v$ is defined in the equality constraint shown in Eq. (16), which works to minimize the control input, $m_x$, as much as possible.

$$m_x^2 + v^2 - m_{max}^2 = 0 \tag{16}$$

The sign of $v$ cannot be determined because of the equality constraint's characteristic. Thus, we introduced a negative sign preceding $R_2$ in Eq. (15) to keep v positive, which leads to a reduction in the value of the cost function $J$.

### D. Weight Functions, $Q, R_1, R_2$

As for the weight functions, $Q, R_1, R_2$, we make their value switched, depending on angular velocities to stabilize all 3 axes, including *x*-axis, which is especially difficult to detumble as shown in Eq. (4). In the beginning of the detumbling process, when the angular rates are relatively large, the control inputs attenuate the angular rates with a large *x*-axis weight function, $Q_x$. On the other hand, at the end of the process, by making other axes' weight functions large, which is normal to the *x*-axis, we can achieve 3-axis detumbling. The specific values and the conditions to switch the weight functions are given in Tables 1, 2, and 3.

Table 1 Definition of the switch conditions

| | |
|---|---|
| Condition 1 | $\omega_x > 0.1$ [deg/s] |
| Condition 2 | $\omega_x < 0.1$ [deg/s] |

Table 2 $Q_x, Q_y, Q_z$

| Weight functions | Condition 1 | Condition 2 |
|---|---|---|
| $Q_x$ | $10^{3.5}$ | $10^{3.5}$ |
| $Q_y$ | $10^{-2}$ | $10$ |
| $Q_z$ | $10^{-2}$ | $10$ |

Table 3 $R_1, R_2$ (constant)

| Weight functions | Value |
|---|---|
| $R_1$ | $10^{-2}$ |
| $R_2$ | $10^{-5}$ |

### E. Update Rule on MPC

There are two approaches to find the optimal input on MPC: to update either Lagrange multiplier or the optimal input matrix defined as Eq. (17). The approach using Lagrange multiplier has a tendency to be unstable; hence, this paper presents the method which renews Eq. (17) at every sampling time. However, the calculation generally takes a large amount of time, so the method updating Eq. (17) is not suitable for the detumbling of small satellites [15].

$$U(t) = \begin{bmatrix} u_0^*(t) \\ \rho_0^*(t) \\ \vdots \\ u_{N-1}^*(t) \\ \rho_{N-1}^*(t) \end{bmatrix} \quad \text{where } u = \begin{bmatrix} m_x \\ v \end{bmatrix} \quad (17)$$

Instead of updating Eq. (17) at every time step, by introducing $F(U(t), \omega(t), t)$ defined in Eq. (18) [16], we find the optimal input matrix $U(t)$ by solving Eq. (18) where $H$ is Hamiltonian defined in Eq. (19), which always has to be zero. The equation to solve finally ends up in an algebraic equation—Eq. (18)—. We can solve Eq. (18) by using an iterative method such as Newton Method; however, to find a solution in less calculation time, we need to evaluate $U(t)$ with another approach; we use a method to update $U(t)$ by using value of $U(t)$ on a preceding time step. To renew $U(t)$ at every time step, we need Eqs. (20) and (21) as constraints because when Eq. (20) holds for every $t$ and Eq. (21) is true at the initial time, we can assume Eq. (18) always holds. Even if $\boldsymbol{F}(\boldsymbol{U}(t), \boldsymbol{\omega}(t), t)$ temporarily becomes a non-zero value, $\boldsymbol{F}(\boldsymbol{U}(t), \boldsymbol{\omega}(t), t)$ converges to zero exponentially.

$$\boldsymbol{F}(\boldsymbol{U}(t), \boldsymbol{\omega}(t), t) \equiv \begin{bmatrix} \left(\frac{\partial H}{\partial u}\right)^T (\boldsymbol{\omega}_0^*(t), u_0^*(t), \lambda_1^*(t), t) \\ m_{x0}^2 + v_0^2 - m_{max}^2 \\ \vdots \\ \left(\frac{\partial H}{\partial u}\right)^T (\boldsymbol{\omega}_{N-1}^*(t), u_{N-1}^*(t), \lambda_N^*(t), t+T) \\ m_{xN-1}^2 + v_{N-1}^2 - m_{max}^2 \end{bmatrix} = \boldsymbol{0} \tag{18}$$

$$H = \frac{1}{2}\{Q_1\omega_x^2 + Q_2\omega_y^2 + Q_3\omega_z^2) + R_1 m_x^2\} - R_2 v$$

$$+\lambda_x \left(\frac{1}{J_x}(J_y - J_z)\omega_y\omega_z\right)$$

$$+\lambda_y \left(\frac{1}{J_y}\{(J_z - J_x)\omega_z\omega_x - B_z m_x\}\right)$$

$$+\lambda_z \left(\frac{1}{J_z}\{(J_x - J_y)\omega_x\omega_y + B_y m_x\}\right)$$

$$+\rho(m_x^2 + v^2 - m_{max}^2) \tag{19}$$

$$\frac{d}{dt}\boldsymbol{F}(\boldsymbol{U}(t), \boldsymbol{\omega}(t), t) = -\zeta \boldsymbol{F}(\boldsymbol{U}(t), \boldsymbol{\omega}(t), t) \quad (\zeta > 0) \tag{20}$$

$$\boldsymbol{F}(\boldsymbol{U}(0), \boldsymbol{\omega}(0), 0) = \boldsymbol{0} \tag{21}$$

To solve Eq. (20) at every time step, Eq. (20) needs to be simplified as follows [17, 18]. (parameters are abbreviated.)

$$\frac{\partial \boldsymbol{F}}{\partial \boldsymbol{U}} \dot{\boldsymbol{U}}(t) = -\zeta \boldsymbol{F} - \frac{\partial \boldsymbol{F}}{\partial \boldsymbol{U}} \dot{\boldsymbol{\omega}}(t) - \frac{\partial \boldsymbol{F}}{\partial t} \tag{22}$$

Equation (22) practically can be approximated as follows. The value $h$ is a constant and small enough.

$$\frac{F(U + h\dot{U}, \omega + h\dot{\omega}, t + h) - F(U, \omega + h\dot{\omega}, t + h)}{h}$$
$$= -\zeta F - \frac{F(U + h\dot{U}, \omega + h\dot{\omega}, t + h) - F(U, \omega, t)}{h} \quad (23)$$

Applying Generalized Minimal Residual (GMRES) method to Eq. (23) [19], we can calculate $\dot{U}$, from which we can update a new control input matrix $U$. Moreover, with the GMRES method, we can calculate values in constant time. This is the way to update the new control input at every time step.

## IV. Simulation Results

This section illustrates the simulation results of an attempt to detumble a satellite by implementing both B-dot control algorithm and MPC on a 600 km altitude sun-synchronous orbit. As shown in Tables 4, 5, 6, and 7, moments of inertia of the Q-Li satellite, maximal magnitude of the magnetic torquer, initial angular velocities, and six orbital elements where the Q-Li satellite are deployed are all provided.

Table 4 Moment of inertia of the Q-Li satellite

| Moment of inertia | Value   [kg/m$^2$] |
|---|---|
| $J_x$ | 0.0045870 |
| $J_y$ | 0.031420 |
| $J_z$ | 0.031249 |

Table 5 Maximum magnitude of the magnetic torquer

| $m_{max}$ | 10.0  [A·m$^2$] |
|---|---|

Table 6 Initial angular velocities

| Initial angular velocities | Value |
|---|---|
| $\omega_x, \omega_y, \omega_z$ | 0.100 [rad/s] (=5.73 [deg/s]) |

Table 7 Six elements of the orbit

| Semi-major axis | 6691.6 [km] |
|---|---|
| Eccentricity | 0.00046440 |
| Inclination | 96.700[deg] |
| Right Ascension of Ascending Node | 100.90 [deg] |
| Argument of perigee | 119.70 [deg] |
| Mean anomaly | 240.49 [deg] |

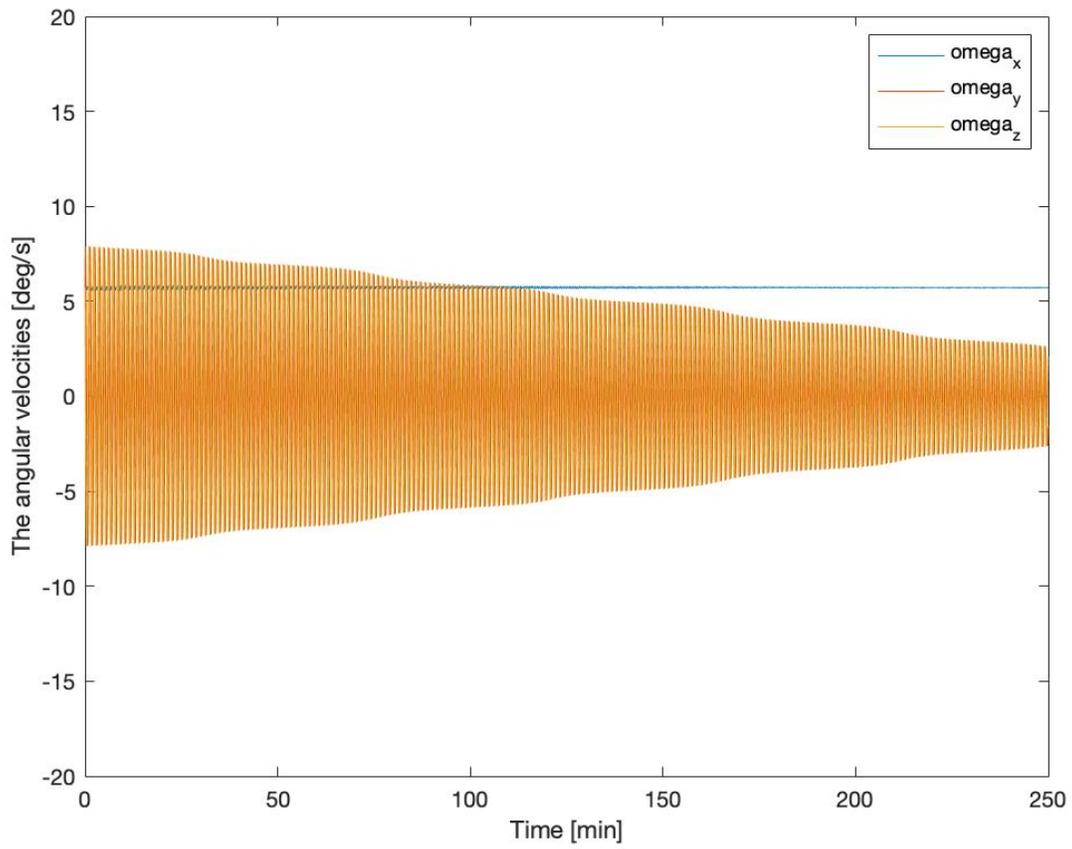

**Fig. 1 Time history of angular velocities with B-dot algorithm**

Figure 1 demonstrates that the B-dot control algorithm has, as we discussed above, difficulty in stabilizing *x*-axis. Moreover, even though the torquer decreases *y* and *z*-axis angular velocities, they are not completely converged to zero in a set time frame: 250 [min].

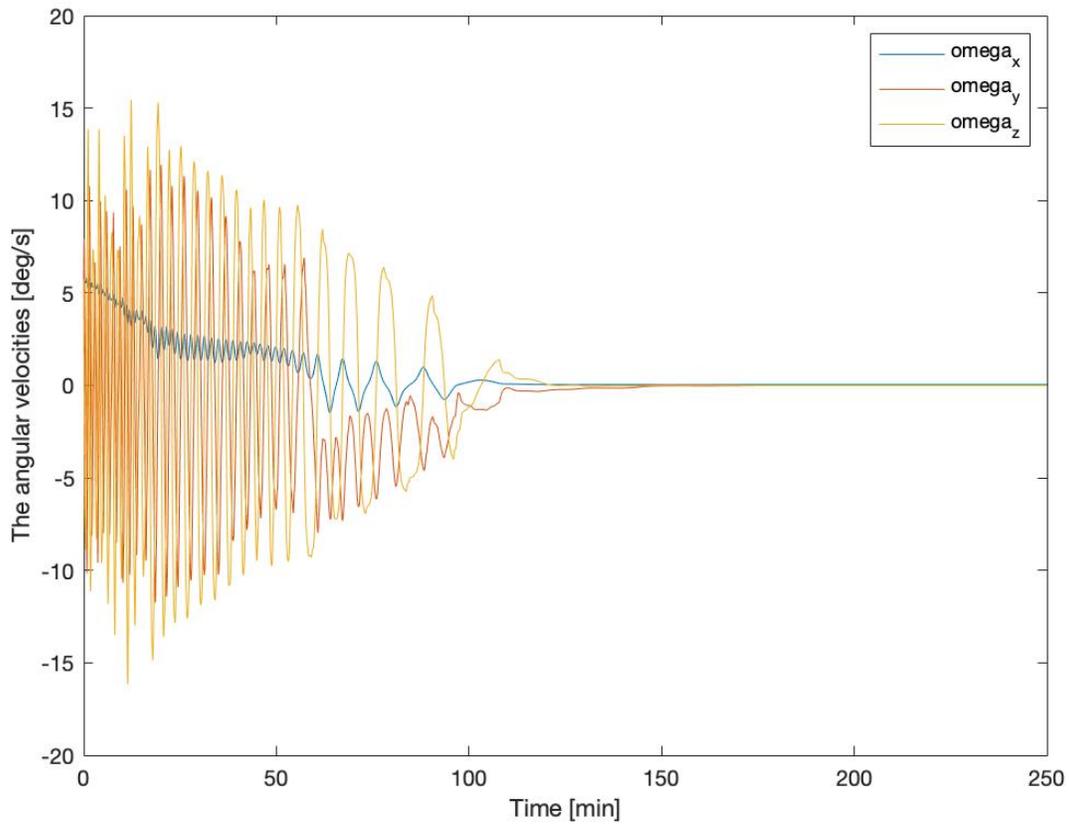

**Fig. 2 Time history of the angular velocities with MPC**

Figure 2 shows the MPC detumbles all three axes in a smaller amount of time. As Figure 1 demonstrates, in general, it is challenging to stabilize *x*-axis angular rate; however, the stabilization on the *x*-axis is satisfied as well as the other axes. Intriguingly, the angular velocities around *y* and *z*-axis get larger than that of the beginning of the detumbling process, which cannot be seen in any time of the process with B-dot algorithm. This is triggered by the difference between the weights of MPC weight functions in the cost function. To preferentially stabilize the *x*-axis angular rate, MPC increases other axes' angular rates. That is, MPC first focuses on detumbling around the *x*-axis, which results in amplifying the angular velocities on the other perpendicular axes in the beginning of the process.

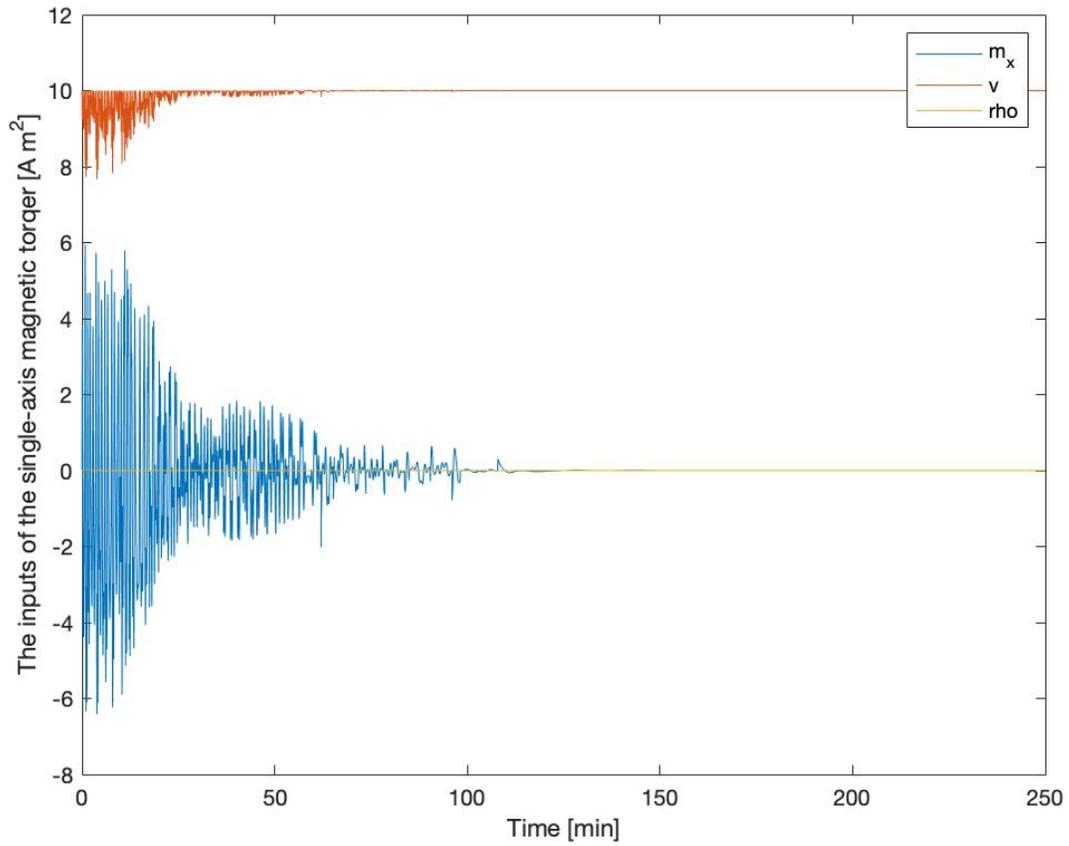

**Fig. 3 Time history of the MPC inputs of the satellite**

Figure 3 gives time history of the control inputs. The dummy input $v$ is always greater than 0 because of the negative sign preceding $R_2$.

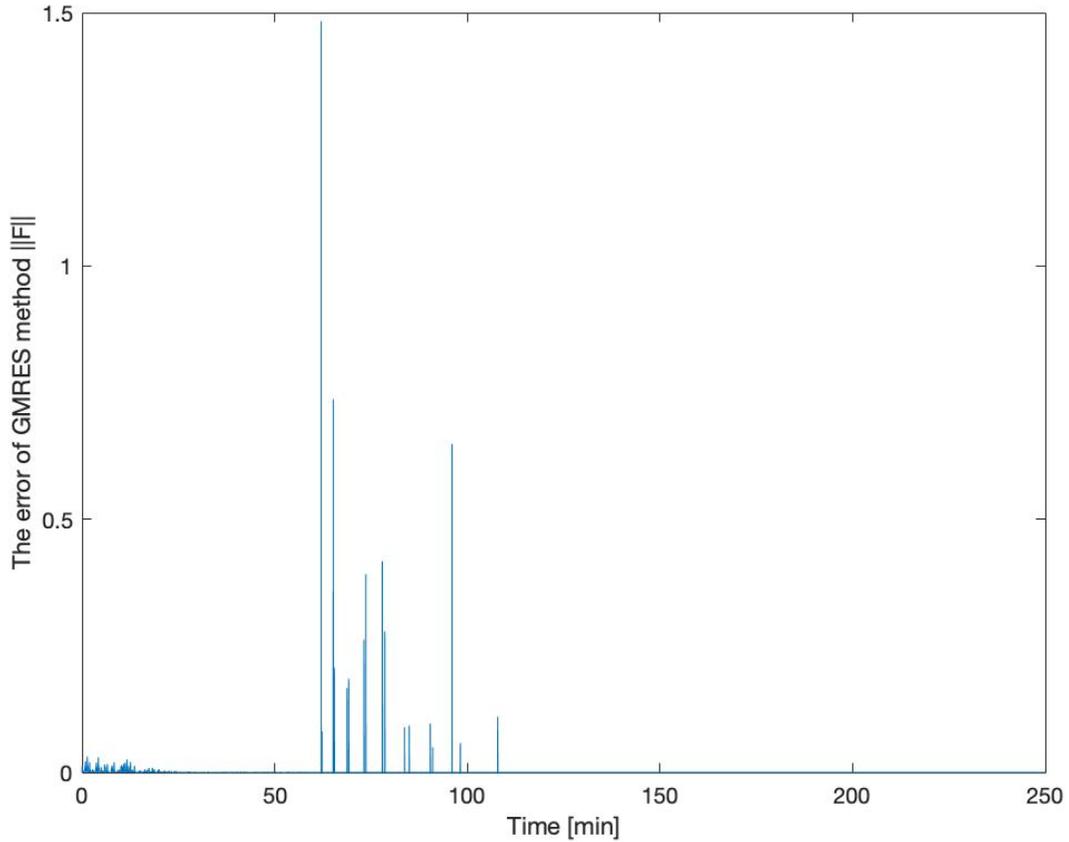

**Fig. 4 The error of GMRES method**

The error of the GMRES method is given as norm of *F* matrix as shown in Fig. 4. The maximum error is less than 1.5, which indicates the GMRES method is completed with a sufficient degree of accuracy, and the control input using MPC satisfies the necessary conditions for optimality.

## V.  Conclusions

To detumble 3-axis angular rates with a single axis magnetorquer, we show the applicability of MPC underactuated controller. Adjusting weight functions with respect to angular rates, MPC stabilizes *x*-axis, where an input torque is not generated. The difficulty of an underactuated B-dot algorithm system is also shown both theoretically and as in the simulation results. The detailed analyses of B-dot control and the model predictive control on the application possibility for the Q-Li satellite is also demonstrated.